\documentstyle[12pt]{article}

\newcommand \beq{\begin{eqnarray}}
\newcommand \eeq{\end{eqnarray}}

\catcode`\@=11

\def\ps@myheadings{\let\@mkboth\@gobbletwo
\def\@oddhead{\hbox{} 
\rightmark\hfil\ninerm\thepage}
\def\@oddfoot{}\def\@evenhead{\ninerm\thepage\hfil 
\leftmark\hbox{}}\def\@evenfoot{}
\def\sectionmark##1{}\def\subsectionmark##1{}}

\evensidemargin
0.30truein
\def\abstracts#1{{
        \centering{\begin{minipage}{30pc}\baselineskip=22pt\noindent
        \centerline{\tenrm ABSTRACT}\vspace{0.3cm}
        \parindent=0pt #1
        \end{minipage} }\par}}
\begin{document}

\begin{titlepage}
\begin{flushright}

{Saclay-T94/02}
\end{flushright}
\vspace*{3cm}
\begin{center}
\baselineskip=13pt
{\Large NON-ABELIAN EXCITATIONS OF THE\\QUARK-GLUON PLASMA\\}
\vskip0.5cm
Jean-Paul BLAIZOT\footnote{CNRS}  and
Edmond IANCU\\
{\it Service de Physique Th\'eorique\footnote{Laboratoire de la Direction des
Sciences de la Mati\`ere du Commissariat \`a l'Energie
Atomique}, CE-Saclay \\ 91191 Gif-sur-Yvette, France}\\
\vskip0.5cm
January 1994
\end{center}

\vskip 2cm
\abstracts{We present new, non-abelian, solutions to the equations of
motion which describe the collective excitations of a quark-gluon plasma
at high temperature. These solutions correspond to spatially uniform
color oscillations.
}
\vskip 4cm

\begin{flushleft}
Submitted to Physical Review Letters
\end{flushleft}

\end{titlepage}


\def\square{\hbox{{$\sqcup$}\llap{$\sqcap$}}}   
\def\grad{\nabla}                               
\def\del{\partial}                              

\def\frac#1#2{{#1 \over #2}}
\def\smallfrac#1#2{{\scriptstyle {#1 \over #2}}}
\def\half{\ifinner {\scriptstyle {1 \over 2}}
   \else {1 \over 2} \fi}

\def\bra#1{\langle#1\vert}              
\def\ket#1{\vert#1\rangle}              

\def\simge{\mathrel{%
   \rlap{\raise 0.511ex \hbox{$>$}}{\lower 0.511ex \hbox{$\sim$}}}}
\def\simle{\mathrel{
   \rlap{\raise 0.511ex \hbox{$<$}}{\lower 0.511ex \hbox{$\sim$}}}}


\def\parenbar#1{{\null\!                        
   \mathop#1\limits^{\hbox{\fiverm (--)}}       
   \!\null}}                                    
\def\nunubar{\parenbar{\nu}}
\def\ppbar{\parenbar{p}}


\def\buildchar#1#2#3{{\null\!                   
   \mathop#1\limits^{#2}_{#3}                   
   \!\null}}                                    
\def\overcirc#1{\buildchar{#1}{\circ}{}}


\def\slashchar#1{\setbox0=\hbox{$#1$}           
   \dimen0=\wd0                                 
   \setbox1=\hbox{/} \dimen1=\wd1               
   \ifdim\dimen0>\dimen1                        
      \rlap{\hbox to \dimen0{\hfil/\hfil}}      
      #1                                        
   \else                                        
      \rlap{\hbox to \dimen1{\hfil$#1$\hfil}}   
      /                                         
   \fi}                                         %


\def\subrightarrow#1{
  \setbox0=\hbox{
    $\displaystyle\mathop{}
    \limits_{#1}$}
  \dimen0=\wd0
  \advance \dimen0 by .5em
  \mathrel{
    \mathop{\hbox to \dimen0{\rightarrowfill}}
       \limits_{#1}}}                           

\def\real{\mathop{\rm Re}\nolimits}     
\def\imag{\mathop{\rm Im}\nolimits}     

\def\tr{\mathop{\rm tr}\nolimits}       
\def\Tr{\mathop{\rm Tr}\nolimits}       
\def\Det{\mathop{\rm Det}\nolimits}     

\def\mod{\mathop{\rm mod}\nolimits}     
\def\wrt{\mathop{\rm wrt}\nolimits}     


\def\TeV{{\rm TeV}}                     
\def\GeV{{\rm GeV}}                     
\def\MeV{{\rm MeV}}                     
\def\KeV{{\rm KeV}}                     
\def\eV{{\rm eV}}                       

\def\mb{{\rm mb}}                       
\def\mub{\hbox{$\mu$b}}                 
\def\nb{{\rm nb}}                       
\def\pb{{\rm pb}}                       

%
\def\journal#1#2#3#4{\ {#1}{\bf #2} ({#3})\  {#4}}

\def\AdvPhys{\journal{Adv.\ Phys.}}
\def\AnnPhys{\journal{Ann.\ Phys.}}
\def\EurophysLett{\journal{Europhys.\ Lett.}}
\def\JApplPhys{\journal{J.\ Appl.\ Phys.}}
\def\JMathPhys{\journal{J.\ Math.\ Phys.}}
\def\LettNuovoCimento{\journal{Lett.\ Nuovo Cimento}}
\def\Nature{\journal{Nature}}
\def\NPA{\journal{Nucl.\ Phys.\ {\bf A}}}
\def\NPB{\journal{\it {Nucl.\ Phys.\ {\bf B}}}}
\def\NuovoCimento{\journal{Nuovo Cimento}}
\def\Physica{\journal{Physica}}
\def\PLA{\journal{Phys.\ Lett.\ {\bf A}}}
\def\PLB{\journal{Phys.\ Lett.\ {\bf B}}}
\def\PhysRev{\journal{Phys.\ Rev.}}
\def\PRC{\journal{Phys.\ Rev.\ {\bf C}}}
\def\PRD{\journal{\it {Phys.\ Rev.\ {\bf D}}}}
\def\PRL{\journal{\it {Phys.\ Rev.\ Lett.}}}
\def\PhysRept{\journal{Phys.\ Repts.}}
\def\ProcNatlAcadSci{\journal{Proc.\ Natl.\ Acad.\ Sci.}}
\def\ProcRoySoc{\journal{Proc.\ Roy.\ Soc.\ London Ser.\ A}}
\def\RevModPhys{\journal{Rev.\ Mod.\ Phys.}}
\def\Science{\journal{Science}}
\def\SovPhysJETP{\journal{\it {Sov.\ Phys.\ JETP}}}
\def\SovPhysJETPLett{\journal{Sov.\ Phys.\ JETP Lett.}}
\def\SovJNuclPhys{\journal{\it {Sov.\ J.\ Nucl.\ Phys.}}}
\def\SovPhysDoklady{\journal{Sov.\ Phys.\ Doklady}}
\def\ZPhys{\journal{Z.\ Phys.}}
\def\ZPhysA{\journal{Z.\ Phys.\ A}}
\def\ZPhysB{\journal{Z.\ Phys.\ B}}
\def\ZPhysC{\journal{Z.\ Phys.\ C}}

\baselineskip=22pt
\setcounter{equation}{0}
\parindent=20pt

The long wavelength excitations of a quark-gluon plasma  are
 collective excitations which are described by nonlinear equations
generalizing the classical Yang-Mills equations in the vacuum.
Most studies of such equations have been limited so far to their
weak field limit, where the modes reduce essentially to abelian-like
plasma waves. The purpose of this letter is to present new, truly
non-abelian, solutions that we have obtained recently.
At leading order in the gauge coupling
$g$, (we assume $g\ll 1$ in the high temperature deconfined plasma),
the collective dynamics is entirely described by a set of effective equations
for the soft gauge mean fields $A^\mu_a(x)$ which describe the long wavelength
($\lambda\sim 1/gT$) and low frequency ($\omega\sim gT$) excitations
 ( $T$ denotes the temperature)\cite{us,Pisarski}.
(Throughout this work, the  greek indices refer to Minkovski space, while the
latin subscripts are color indices for the adjoint representation of the
gauge group $SU(N)$).  The equations satisfied by  $A^\mu_a(x)$ are
\beq
\label{ava}
\left [\, D^\nu,\, F_{\nu\mu}(x)\,\right ]_a
\,=\,j_{\mu\,a}^{ind}(x),
\eeq
where $D^\mu = \del^\mu+igA^\mu(x),$ ($A^\mu\equiv A^\mu_a T^a$),
and  $F^{\mu\nu}= [D^\mu, D^\nu]/(ig) = F^{\mu\nu}_aT^a$.
The induced current $j_\mu^{ind}$  describes the  response of the plasma to the
color gauge fields $A_a^\mu$. It is proportional to the
 fluctuations in the phase-space color densities of quarks
and gluons. Its expression is\cite{us,EMT}
\beq\label{j1}
j^{ind}_{\mu\,a}(x)\,=\,3\,\omega^2_p\int\frac{d\Omega}{4\pi}
\,v_\mu\,W_a(x;v),\eeq
for retarded boundary conditions ($A_a^\mu(x)\to 0$ as $x_0\to -\infty$).
The notations here are as follows:
$\omega^2_p\equiv (2N+N_{\rm f})g^2 T^2/18$  is the {\it plasma frequency},
 $v^\mu\equiv (1,\,{\bf v})$, where ${\bf v}\equiv
{\bf k}/k$ is the velocity of the hard particle with momentum ${\bf k}$
($k\equiv |{\bf k}|$), and the  integral $\int
d\Omega$ runs over all the directions of the unit vector ${\bf v}$.
Furthermore, the functions $W_a(x;v)$ are generally nonlocal and nonlinear
functionals of the gauge fields, defined by
\beq\label{W}
W_a(x;v)\,=\, \int_0^\infty du\, U_{ab}(x,x-vu)\, {\bf v}\cdot {\bf E}_b(x-vu)
\eeq
where $E^i\equiv F^{i0}$ is the chromoelectric field,  and
  $U(x,y)$  is the parallel transporter along the straight line
$\gamma$   joining $x$ and $y$.
The induced current (\ref{j1}) is covariantly conserved,
$\left[D^\mu,\,j_\mu^{ind}(x)\right]=0$.
The energy density of an arbitrary gauge field configuration in the plasma
has been recently computed as\cite{EMT}
\beq\label{enden}
T^{00}(x)\,=\,\frac{1}{2}\Bigl({\bf E}_a(x)\cdot{\bf E}_a(x)\,+\,
{\bf B}_a(x)\cdot{\bf B}_a(x)\Bigr)\,+\,\frac{3}{2}\,
\omega^2_p\int\frac{d\Omega}{4\pi}\,W_a^0(x;v)\,W_a^0(x;v),\eeq
with $B^i_a(x)\equiv -(1/2)\epsilon^{ijk}\,F^{jk}_a(x)$.
Remark that $T^{00}(x)$ is manifestly positive,
and so is therefore the excitation energy ${\cal E}(t)
=\int d^3x\,T^{00}(t,{\bf x})$, at any time $t$.

The solutions to the non-abelian field equations (\ref{ava})
have direct physical relevance: they
correspond to collective excitations of the high temperature quark-gluon
plasma. In this letter, we study particular solutions
  which are uniform in  space.
(Another interesting limit, that of a static field configuration,
has been considered recently\cite{Liu93}, with the conclusion
that no finite energy solution exists).
For convenience, we choose the temporal gauge, $A^0_a(x)\,=\,0$. Hence,
$A^\mu_a(x)\equiv (0, {\bf A}_a(t))$.

For uniform fields, the functions
  $W_a({t};v)$ are simple local functionals of the gauge
potentials\cite{EMT},
\beq
W_a({t};v)&=&-\,{\bf v}\cdot {\bf A}_a({t}),\eeq
and the same holds for the induced current
(\ref{j1}), ${\bf j}^{ind}_a(t)
\,=\,-\omega_p^2\,{\bf A}_a(t)$, $\rho^{ind}_a(t)\,=\,0$,
and for the energy density (\ref{enden}),
\beq\label{ent}
T^{00}(t)&=&\frac{1}{2}\biggl(\frac{d{\bf A}_a}{dt}\cdot
\frac{d{\bf A}_a}{dt}\,+\,\omega_p^2\,{\bf A}_a\cdot {\bf A}_a\biggr)\,+\,
\frac{g^2}{4}\,f^{abc}\,f^{ade}\Bigl({\bf A}_b\cdot {\bf A}_d\Bigr)\,
\Bigl({\bf A}_c\cdot {\bf A}_e\Bigr).\eeq
We have used here the expressions of the field strengths in terms of the
 vector potentials,
 \beq\label{EBt}
{\bf E}_a(t)\,=\,-\,\frac{d{\bf A}_a}{dt},\qquad
{\bf B}_a(t)\,=\,\frac{g}{2}\,f^{abc}\,{\bf A}_b(t)\times {\bf A}_c(t).\eeq

The field equations (\ref{ava}) reduce then to
 the following equations for the vector potentials ${A}^i(t)=A^i_a(
t)\,T^a$,
\beq\label{jet}
\frac{d^2 A^i}{d t^2}\,+\,\omega^2_p A^i\,+\,g^2\,\biggl [\,
\Bigl[ A^i, A^j\Bigr],\,A^j\,\biggr]\,=\,0,\eeq
together with a  constraint (Gauss's law)
\beq\label{rhot}
\biggl[A^i,\frac{d A^i}{dt}\biggr]\,=\,0,\eeq
corresponding to the component  $\mu=0$  of eq.~(\ref{ava}).
These equations are similar to the
 classical Yang-Mills equations  in  the vacuum, which have been
extensively investigated already for the case of
 $SU(2)$\cite{Baseyan79,Chirikov81,Shur82,Chang83}.
They differ, however, by the presence of the thermal mass term $\omega_p^2A^i$,
 which, as we shall see, has a strong effect on the dynamics.
Note incidentally that, for gauge fields satisfying  Gauss's law (\ref{rhot}),
 the Poynting vector $S^i\equiv T^{i0}$ vanishes\cite{EMT}.

The constraint (\ref{rhot}) is satisfied, in particular, by field
configurations of the form
$A^i_a(t)\,=\,{\cal A}^i_a\,h^i(t)$ (no summation over $i$),
 with constant ${\cal A}^i_a$ and arbitrary  functions $h^i(t)$.
Indeed, for such fields, the  three color vectors $\{A^i_a(t)\}$
and $\{dA^i_a/dt\}$ ($i=1,\,2,\,3$) are  parallel in color space.
In the rest of this letter we restrict ourselves to
  $SU(2)$, and assume ${\cal A}^i_a=\delta^i_a$. This Ansatz is equivalent
to that proposed by Baseyan et al.\cite{Baseyan79}, up to
a trivial global gauge rotation \cite{Chang83}.
The functions $h_i(t)$ then satisfy
\beq\label{h1}
\frac{d^2 h_1}{d t^2}\,+\,\omega^2_p\, h_1\,+\,g^2\,h_1\,\bigl(
h_2^2+h_3^2\bigr)\,=\,0,\eeq
plus two similar equations for $h_2$ and  $h_3$.
The associated  energy density,
\beq\label{t00}
T^{00}&=&\frac{1}{2}\sum_i\left(\left(\frac{dh_i}{dt}\right )^2+
\omega_p^2\,h_i^2\right)\,+\,
\frac{g^2}{2}\,\Bigl(h_1^2\,h_2^2\,+\,h_1^2\,h_3^2\,
+h_2^2\,h_3^2\Bigr),\eeq
is an integral of motion and acts as an effective Hamiltonian for the
functions $h_i(t)$.

At this point, it is convenient to make a scale transformation and define
the dimensionless variable $x\equiv \omega_p t$, as well as the dimensionless
functions $f_i(x)\equiv (g/\omega_p)\,h_i(t)$. We also
assume $f_3=0$, with no significant loss of generality.
We obtain then  for $f_1$ and $f_2$ the coupled nonlinear equations
\beq\label{fsys}
{\ddot f}_1(x)\,+\,\Bigl[1+\Bigl(f_2(x)\Bigr)^2\,\Bigr]\,f_1(x)&=&0,\nonumber
\\{\ddot f}_2(x)\,+\,\Bigl[1+\Bigl(f_1(x)\Bigr)^2\,\Bigr]\,f_2(x)&=&0,\eeq
where the overdots indicate derivatives with respect to $x$. The energy
density (\ref{t00}) is then
\beq\label{T00}
T^{00}&=&\frac{\omega_p^4}{g^2}\,\frac{1}{2}\,\biggl(
{\dot f}_1^2+{\dot f}_2^2 + f_1^2+f_2^2 +f_1^2\,f_2^2\biggr)\equiv
\frac{\omega_p^4}{g^2}\,{\cal H}.\eeq
The Hamiltonian  ${\cal H}$ is that of a system of two nonlinearly
coupled harmonic oscillators with coordinates $f_1$ and $f_2$.
Note that the different parameters characterizing the initial system
 (i.e., $\omega_p$, $g$, $T^{00}$) combine into a single, dimensionless, one,
$\theta^2\equiv (g^2/\omega_p^4)T^{00}$, which measures the total energy
of the mechanical system: ${\cal H}=\theta^2$.

In deriving eq.(\ref{ava}), we have assumed the gauge fields
 to be weak ($A\simle T$) and slowly varying ($\del A\sim gTA$)\cite{us}.
Remembering that $\omega_p\sim gT$, one sees that these conditions imply
  $|f_i(x)|\simle 1$ and $|{\dot f}_i(x)|
\sim |f_i(x)|$, so that $\theta$ is, at most, of  order unity,
$\theta\simle 1$. These limitations are consistent with the dynamics
described by eqs.~(\ref{fsys}). The quadratic  terms $f_1^2+f_2^2$ in the
Hamiltonian (\ref{T00}), which originates from the thermal mass,
play an essential role in this respect.
Because of them, and of energy conservation,
a trajectory $\{f_i(x)\}$ cannot
leave the {\it bounded} domain delineated by the equipotential
lines $f_1^2+f_2^2+f_1^2\,f_2^2\,=\,2\theta^2$ (Fig.1). Let us assume
$\theta\simle 1$. Then, since  $|f_i(x)|\le \sqrt 2\,\theta$ (see Fig.1),
it follows that $|f_i(x)|\simle 1$ for any $x$.
Furthermore, the quantity $(1+f_i^2(x))$, which plays the role of an effective
frequency squared for the motion in the direction $j\not = i$,
 remains of order $1$ for any $x$,
so that $f_i$ and ${\dot f}_i$ remain of the same order of magnitude.
Consequently, if the conditions on the functions $f_i$ mentioned above are
satisfied for some $x_0$, then they are valid for any $x$.
 The situation here
is different from the classical, vacuum, case, were the equipotential lines
 are given by  $h_1^2\,h_2^2\,=\,(2/g^2)T^{00}$; then, for given energy
$T^{00}$,  the motion could extend
arbitrary far  along the $h_1$ or the $h_2$ axis.
Using the language of dynamical systems, one may observe that,
 for the system (\ref{fsys}), the
origin of the four-dimensional phase-space $(f_1, {\dot f}_1, {f}_2,
{\dot f}_2)$ is an {\it elliptic fixed point}, corresponding to a neutrally
 stable equilibrium\cite{Arnold78}.
 At $T=0$, this same point is marginally unstable. Thus, apart from ensuring
bounded trajectories $\{f_i(x)\}$, the quadratic
 terms $f_i^2$ in eq.~(\ref{T00})
also improve the stability of the system.

Let us consider now simple, periodic, solutions to the system (\ref{fsys}).

(a) The simplest motion is one-dimensional  along
 the axes 1 or 2. Assume, e.g.,  $f_2=0$. We get then
a simple harmonic oscillator equation for  $f_1$,
\beq
{\ddot f}_1(x)\,+\,f_1(x)&=&0.\eeq
The general solution is  $f_1\,=\,a_1\,\cos x+a_2\,\sin x$, with
 $a_1^2 + a_2^2= 2\theta^2$. The corresponding
field configuration, $A^1(t)\,=\,
(C_1\cos\omega_pt\,+\,C_2\sin\omega_pt)\,T^1$,
 (with $C_i\equiv (\omega_p/g)a_i$), and $A^2=A^3=0$,
describes periodic oscillations (with period ${\cal T}_0=2\pi/\omega_p$)
in space-color direction 1.

(b) Another one-dimensional, but less trivial, example  corresponds to
periodic solutions with $f_2(x)=\pm f_1(x)\equiv \pm f(x)$.
These  solutions describe in or out of phase
oscillations of the  colors 1 and 2.
The function $f(x)$ satisfies the nonlinear equation
\beq\label{fnon}
{\ddot f}(x)\,+\,f(x)\,+\,f^3(x)&=&0,\eeq
for which the integral of motion is \beq
\theta^2\,=\,{\dot f}^2\,+\,f^2\,+\,\frac{1}{2}\,f^4.\eeq
A particular solution to (\ref{fnon}) is
\beq\label{fsol}
f(x)\,=\,f_\theta\,{\rm cn}\Bigl((2\theta^2+1)^{1/4}(x-x_0);\,k\Bigr),\eeq
where ${\rm cn}(x;k)$ is the Jacobi elliptic cosine of argument $x$ and
modulus $k$, and $x_0$ is the arbitrary origin of the time. The
parameters $k$ and $f_\theta$ are related
to $\theta$  by
\beq\label{k}
k\,=\,\frac{1}{\sqrt 2}\,\biggl(1\,-\,\frac{1}{\sqrt {2\theta^2+1}}\biggr)^
{1/2},\eeq
and
\beq\label{ftheta}
f_\theta\,=\,\Bigl({\sqrt {2\theta^2+1}}\,-\,1\Bigr)^{1/2}.\eeq
It can be easily seen that $f_1=\pm f_2=\pm f_\theta$  are
the coordinates of the intersection points between the trajectories
and the equipotential lines in Fig. 1.
The solution (\ref{fsol}) corresponds to the initial conditions
$f(x_0)=f_\theta$ and ${\dot f}(x_0)=0$. For the in-phase oscillations, the
 associated gauge potentials are $A^1(t)\,=\, h(t)\, T^1$,  $A^2(t)\,=\,
h(t)\, T^2$,  and $A^3(t)\,=\, 0$, where
$h(t)\,=\,(\omega_p/g)\,f(\omega_p t)$
is a periodic function, with period
\beq\label{T}
{\cal T}\,=\,\frac{4}{\omega_p}\,\frac{1}{(2\theta^2+1)^{1/4}}\,K(k),\eeq
and $K(k)$ is the complete elliptic integral of modulus k. Since
$\theta\simle 1$,  ${\cal T}$  remains of order of
 ${\cal T}_0=2\pi/\omega_p$. Thus, it is the plasma frequency
$\omega_p\sim gT$ which controls the time variation of the nonlinear color
oscillations (\ref{fsol}), for any value of the parameter $\theta\simle 1$.
This is true, in particular, for {\it small oscillations}, that is, when
the energy density $T^{00}\to 0$, so that  $\theta\approx
f_\theta\ll 1$ and $k\to 0$. Then, the periodic solution (\ref{fsol}) reduces
to a simple harmonic oscillation, and ${\cal T}\to {\cal T}_0$.
In contrast, in the zero temperature
case, as $T^{00}\to 0$, the corresponding period is diverging\cite
{Baseyan79}. (The $T=0$ expression for ${\cal T}$ is obtained
 from eq.~(\ref{T}) by replacing
 $(2\theta^2+1)^{1/4}\omega_p\rightarrow (2g^2T^{00})^{1/4}$ and
 $k\rightarrow 1/{\sqrt 2}$.)

For this solution, the non-trivial field strengths
are $E^i_a=-\delta^i_a\,{\dot h}$, with $i=1,\,2$,
and $B^3_a=g\,h^2\,\delta^3_a$. Accordingly, the three vectors ${\bf E}_1$,
${\bf E}_2$, and ${\bf B}_3$ are mutually orthogonal.

An important question
concerns the stability  of the periodic orbits of the system (\ref{fsys}).
For the corresponding equations at zero temperature, it has been
 shown, through a numerical  analysis,
 that unstable periodic trajectories
exist\cite{Baseyan79}. In particular, the in-phase oscillations of two colors
 (the vacuum analog of (\ref{fsol})) are unstable\cite{Chirikov81,Shur82}.
In the case of the system (\ref{fsys}), the presence of  the linear terms
$f_1$ and $f_2$, which originate from the thermal mass in (\ref{t00}),
guarantees the stability of the solutions, for small oscillations and for
 most initial conditions. This is verified in the numerical analysis of the
system (\ref{fsys}) performed in Ref.~\cite{Matinyan81} in the
 physically different context of a classical Yang-Mills-Higgs
system in the vacuum. There, the mass terms for the gauge fields are generated
through spontaneous symmetry breaking and
 are proportional to the vacuum expectation value of the scalar
field, $\eta=\langle \phi\rangle$. The correspondence between the two
sets of equations is simply given by  $\omega_p^2\rightarrow g^2\eta^2/2$.
 The numerical experiment in Ref.~\cite{Matinyan81} clearly shows that
for $\theta\ll 1$ (e.g., for $\theta\approx 0.2$) most of the trajectories
are quasi-periodic.
As $\theta$ increases,  the stochastic motion strongly develops
 and, beyond a critical value  $\theta^2_c\approx 6.6$, it fills
 the entire permissible range of motion in the phase space.
As the above value
of  $\theta_c$ stays in the limits of validity of the present approach, it
would be worth to further investigate the physical signification of this
transition. Analogous findings, concerning the non-integrability and the
stochastic behaviour of the dynamical system (\ref{fsys}), are presented
in Refs. \cite{Pullen} and \cite{Meyer}.

In conclusion, we have studied here global color oscillations of the high
temperature quark-gluon plasma, by looking at the spatially uniform limit
of the effective equations of motion  derived
in Refs.~\cite{us}. We have shown that the plasma frequency
$\omega_p=gT(N+N_f/2)^{1/2}/3$ characterizes the inertial properties of the
plasma  both for the abelian-like and for the genuine non-abelian collective
motion. The phase-space motion is bounded and quasi-periodic for small
oscillations and for most initial conditions. We have obtained explicit
nonlinear solutions for $SU(2)$ which describe in or out of phase
 oscillations of two colors (the generalization
to the in-phase oscillations of the three colors, i.e., $h_1=h_2=h_3$ in
eq.~(\ref{h1}), is straightforward\cite{EMT}).
Such solutions can be imbedded in any
larger $SU(N)$ theory in the standard way\cite{Actor}.

\vspace{1cm}
\noindent
{\bf Acknowledgements}\\
\noindent We would like to thank D. Ullmo far having indicated to us Refs.
\cite{Pullen} and \cite{Meyer}.
\newpage




\end{document}